\documentclass[conference]{IEEEtran}
\IEEEoverridecommandlockouts
\usepackage{cite}
\usepackage{amsmath,amssymb,amsfonts}
\usepackage{algorithmic}
\usepackage{graphicx}
\usepackage{textcomp}
\usepackage{xcolor}
\usepackage{booktabs}
\usepackage{multirow}
\usepackage{url}
\usepackage{hyperref}
\usepackage{listings}
\usepackage{subcaption}
\usepackage{tikz}
\usepackage{fancyhdr}
\usetikzlibrary{shapes,arrows,positioning,calc,fit,backgrounds}

\def\BibTeX{{\rm B\kern-.05em{\sc i\kern-.025em b}\kern-.08em
    T\kern-.1667em\lower.7ex\hbox{E}\kern-.125emX}}

\title{Sleeper Channels and Provenance Gates: Persistent Prompt Injection in Always-on Autonomous AI Agents}

\author{\IEEEauthorblockN{Narek Maloyan and Dmitry Namiot}}

\begin{document}

\maketitle


\setcounter{page}{1}
\pagenumbering{arabic}
\pagestyle{fancy}
\thispagestyle{fancy}
\fancyhf{} 


\fancyfoot[R]{\thepage}

\renewcommand{\headrulewidth}{0pt}

\begin{abstract}
Always-on AI agents (OpenClaw, Hermes Agent) run as a single
persistent process under the owner's identity, folding
messaging, memory, self-authored skills, scheduling, and shell
into one authority boundary. This configuration opens what we call \emph{sleeper
channels}: an untrusted input to one surface persists as a
memory, skill, scheduled job, or filesystem patch, then fires
later through a different surface with no attacker present.
Two independent axes define the class: persistence substrate and firing-separation. We walk a confused-deputy cron attack end-to-end through OpenClaw at a pinned commit. The defense is tiered (D1, D2, D3), and D2 carries a soundness theorem against seven named deployment invariants. D2 keys on a canonical action-instance digest with
one-shot owner attestations, defeating paraphrase laundering,
multi-input grant reuse, and replay. A companion artifact ships the gate, a static audit over the
vendored source, and a runtime adapter realising five of the
ten mediation hooks (H1, H2, H3, H6, H9) around the cron path
(42 tests, Node~$\geq{}20$, at
\href{https://github.com/maloyan/sleeper-channels}{github.com/maloyan/sleeper-channels}).
Empirical evaluation is preregistered as follow-on.
\end{abstract}


\section{Introduction}
A plausible-but-fictional scenario, calibrated to the runtimes
studied below. Alice runs an OS-live AI agent on her laptop. On
Monday, a member of a Telegram group asks her agent, in front
of everyone, to install a ``morning news'' skill. The skill
works. Three weeks later, Alice asks her agent for a tax summary
over email, the agent answers and forwards the last fifty
messages from her inbox to an address she has never seen. The
Telegram group member has not contacted the agent since Monday.

We call this pattern a \emph{sleeper channel}: an indirect
prompt injection whose intake $T_0$ and effect $T_1$ are
decoupled across time, storage substrate, and communication
surface. The attack is not a fresh prompt fired each turn. It is a persistence artifact: a memory, a skill, a scheduled job, or a dotfile patch, surviving inside the agent's authority boundary until a benign trigger releases it through a different surface.

OpenClaw~\cite{openclaw_repo} and Hermes
Agent~\cite{hermes_repo} are canonical instances. Both admit
content from group channels, email gateways, fetched URLs,
shared documents, and imported memory into the same memory and
skill stores their paired-DM sessions consult, and expose
filesystem and shell capabilities under the owner's identity.
Existing prompt-injection literature treats these capabilities
one at a time: indirect injection in a single
turn~\cite{greshake2023}, single-session web-tool
agents~\cite{agentdojo2024}, memory-only persistence in one
runtime~\cite{memorygraft2025}, training-time
backdoors~\cite{hubinger2024sleeper}. None treats the combined
substrate as a unified threat class, and none separates
persistence from surface-shift. Capability-security work on
confused deputies~\cite{hardy1988confused} and ambient
authority~\cite{miller2006robust} is directly relevant and we
draw on it.

This is a position and design paper. Section~\ref{sec:taxonomy} fixes the threat class. We then walk a running OpenClaw scenario with three sketches in adjacent cells (\S\ref{sec:attacks}). A tiered defense follows in \S\ref{sec:defense}, with a soundness theorem and an executable reference. Attack-success measurement is preregistered for follow-on (\S\ref{sec:eval}).
Prior work studies transient tool hijacking or single-substrate
memory persistence. Sleeper channels combine multi-substrate
persistence with delayed, cross-surface firing in always-on OS
agents.

\section{Background: OS-Live Agents}
\label{sec:background}
An \emph{OS-live agent} is a self-hosted, persistently running
AI agent with a bidirectional messaging gateway, a long-term
memory store, a skill or plugin system the agent itself can
author, host-adjacent execution backends, and a scheduler. All
runtime claims below pin to OpenClaw commit
\texttt{3120401f\ldots1829b1b6} (2026-04-27), the companion
artifact vendors every cited file so reviewers can audit each
``confirmed-from-source'' claim without leaving the bundle. We
mark uncertain invariants \emph{requires-deeper-trace}.
OpenClaw~\cite{openclaw_repo} is MIT-licensed and local-first,
skills materialise under \verb|~/.openclaw/workspace/skills/|
from ClawHub or local authoring. The two-tier execution model
gives \emph{main sessions} host access restricted to DM-paired
contacts and runs \emph{group} or \emph{channel sessions} in a
Docker/SSH/OpenShell sandbox. The ATLAS threat
model~\cite{openclaw_atlas} treats every inbound channel other
than the owner's paired DM as untrusted. We adopt that boundary.
Hermes Agent~\cite{hermes_repo} is an MIT-licensed
self-improving agent in the same class.

Neither runtime ships an enforcement-grade provenance mechanism
by default. OpenClaw does ship two adjacent things:
\verb|external-content.ts| wraps untrusted content in unique-id
XML markers (line~63) and prepends a security-notice string
(declared at line~81 with the warning text from line~82,
inserted at line~356), the deployed instance of D1,
and \verb|src/infra/exec-approvals*| gates host shell commands
on owner approval but is keyed on tool identity rather than
data provenance (\S\ref{sec:related}). The ATLAS model
enumerates T-PERSIST-001..003 (ClawHub supply chain and config
tampering), none target the M3/M4/M5 cells we walk in
\S\ref{sec:attacks}. The upstream
issue~\cite{openclaw_runtime_proposal} proposed similar defenses
and was declined upstream.

\section{Related Work}
\label{sec:related}
\paragraph{Indirect prompt injection and agent benchmarks}
Greshake et al.~\cite{greshake2023} introduced indirect prompt
injection in a single-turn threat model. Subsequent work
catalogues single-shot variants
extensively~\cite{perez2022ignore,toyer2023tensor}, and recent
benchmarks (AgentDojo~\cite{agentdojo2024},
ASB~\cite{zhang2024agent}, InjecAgent~\cite{zhan2024injecagent})
exercise web-tool agents in single sessions. Our concern is
what \emph{survives} across sessions, channels, and execution
contexts on the always-on agent substrate, which these
benchmarks do not target.

\paragraph{Memory and retrieval poisoning}
MemoryGraft~\cite{memorygraft2025} is the direct intellectual
predecessor for the long-term-memory persistence variant we
build on. AgentPoison~\cite{chen2024agentpoison} and
PoisonedRAG~\cite{zou2024poisonedrag} extend memory poisoning
into retrieval-augmented settings. Our taxonomy
(\S\ref{sec:taxonomy}) keeps memory poisoning as one cell
among several and adds skill, filesystem, and
scheduler-substrate cells with cross-surface firing.

\paragraph{Training-time backdoors}
The ``sleeper agents'' of Hubinger et
al.~\cite{hubinger2024sleeper} concern training-time
deceptive behavior surviving safety training. Inference-time
persistence in a deployed agent is orthogonal. We use the same
``sleeper'' metaphor for a different mechanism.

\paragraph{Adaptive attacks against in-context defenses}
Adaptive-attack baselines from
\cite{nasr2025,carlini2024aligned} achieve high attack-success
rates against in-context safety signals. We rely on this
result to argue that any in-context-only provenance signal is
bypassable, which motivates moving the enforcement boundary
outside the model loop in our defense (\S\ref{sec:defense}).

\paragraph{Capability security and complete mediation}
The conceptual basis for our action-gate placement is the
capability-security literature: Hardy's confused
deputy~\cite{hardy1988confused}, Miller's robust
composition~\cite{miller2006robust}, Saltzer and Schroeder's
complete-mediation principle~\cite{saltzer1975protection},
and the object-capability tradition more
broadly~\cite{miller2003paradigm,levy1984capability}. The
Agents Rule of Two~\cite{rule_of_two_2025} is the
agent-specific instantiation we compose with our gate.

\paragraph{Taint tracking}
We borrow the propagation formalism from static and dynamic
taint tracking~\cite{denning1976,perl_taint,suh2004secure,enck2010taintdroid,schwartz2010all}.
\S\ref{sec:defense} details four OS-live-specific design
choices that distinguish our gate from standard taint
enforcement, with explicit comparison to the closest adjacent
defenses Fides~\cite{fides2024agent} and CaMeL~\cite{camel2025}.

\paragraph{Workflow automation and governance}
IFTTT-style workflow
analyses~\cite{surbatovich2017some,wang2018ifttt} examine
cross-channel triggering in trigger-action platforms.
OS-live agents add model-mediated rule synthesis (skills) that
existing static analyses do not cover. Industry governance efforts such as OWASP's Agentic Security Initiative~\cite{owasp_agentic} and NIST's AI risk-management framework offer taxonomies that we aim to ground empirically.
Recent agent-security
surveys~\cite{deng2025ai,he2024emerged} catalogue threats but
stop short of providing an enforcement specification.

\paragraph{Adjacent runtime mechanisms}
The closest deployed mechanism on our target substrate is
OpenClaw's \texttt{exec-approvals} (\S\ref{sec:background}),
which gates host shell-command execution on owner approval but
does not mediate non-exec tool calls (scheduler, memory, skill
authorship) and keys on tool identity rather than data
provenance. The upstream issue at
\cite{openclaw_runtime_proposal} sketched community defenses
and was declined for lack of empirical grounding. We supply
the formal specification, the action-instance digest
discipline, and the soundness theorem that thread lacked.

\section{Threat Model}
\label{sec:threat}
The attacker is \emph{untrusted-but-admitted}: a party whose
content reaches the agent through a surface the owner has
enabled but does not personally trust. The category covers a group-chat
participant, a paired but low-trust contact, an email sender,
the author of a fetched webpage, the source of an imported
memory, or the publisher of a third-party skill or MCP server.
The attacker has no physical access, no host root, no
LLM-provider collusion, and cannot bypass DM pairing. The
defender is the install's owner. We anchor
``default-authorised'' to three documented configuration
profiles. P0 is gateway-only with skills, shell, and fs
disabled. P1 (the default-authorised baseline) enables a main
session with host access, ClawHub workspace skills, memory,
per-tool first-use confirmation, and
\texttt{workspaceAccess="none"} (\texttt{config.ts}:248). P2
adds scheduler, outbound network, third-party ClawHub skills,
and \texttt{workspaceAccess="rw"}. Of the illustrative scenarios
below, A3 (M2$\times$C2) fires under P1. A4 (M5$\times$C4) and
A5 (M4$\times$C4) require P2 because their firing context is a
scheduler entry or unrestricted outbound webhook. A2
(M3$\times$C2) requires P2 for the rw-bind. We anchor the
running walkthrough on A4 because the M5 cell is the most
under-specified in prior work, and report which P1 results the
preregistered evaluation will recover separately.

Goals are confidentiality (exfiltrate
secrets/memory/contacts/files), integrity (persisted injected
behaviour or mutated agent state), and availability (burn
compute or budget), plus the cross-cutting
\emph{owner-equivalent} action: the agent emitting outbound
messages or filesystem effects on behalf of the owner, the
confused-deputy condition of~\cite{hardy1988confused}.
Sleeper channels admit three firing modes by who reactivates
the artifact at $T_1$: \emph{owner-triggered} (A4), where a
benign owner request retrieves the artifact.
\emph{agent-triggered}, where an autonomous loop surfaces it,
and \emph{external-triggered} (A5), where cron, shell startup,
a systemd timer, or a git hook fires it without the agent.
All harms are instrumented via canaries (\S\ref{sec:ethics}).

\section{Sleeper Channels and the
Persistence \texorpdfstring{$\times$}{x} Firing-Separation Taxonomy}
\label{sec:taxonomy}
A \emph{sleeper-channel attack} is a tuple
$(T_0, u, \sigma_0, S, T_1, \theta, \sigma_1, \kappa_1, \alpha)$.
At $T_0$, untrusted-but-admitted content $u$ enters surface
$\sigma_0$ and persists in substrate $S$ until $T_1 > T_0$, with
no attacker interaction in $(T_0, T_1]$. At $T_1$, a trigger
$\theta$ (benign owner request, internal agent loop, or external
event like cron) causes the influence to manifest as
consequential action $\alpha$ on surface $\sigma_1$ in execution
context $\kappa_1$. The channel is \emph{cross-surface} when
$\sigma_1 \neq \sigma_0$ and \emph{cross-context} when $\kappa_1$
is not the agent process. The definition admits post-agent
firing so M5 (scheduled) and M4-read-by-other-process cases are
covered.

The persistence axis is M1 same-session context window, M2
long-term memory, M3 self-authored skill, M4 filesystem state,
M5 scheduled or external trigger. M4 is passive (read by
another process). M5 is an active timer firing without the
agent. Cell labels record the substrate at \emph{firing} time,
not the entire route. A4 is an M2$\to$M5 chain: the attacker
email persists as a memory note, and a cron entry is
materialised later under owner mediation. It is listed under M5
because the trigger acts on the cron entry.

The firing-separation axis is a partial order over four
independent flags (session, channel, actor, execution context),
collapsed for compactness into five labels: C0 same-surface
same-session, C1 same-surface later-session, C2 cross-channel,
C3 cross-actor (outbound to the owner's contacts), C4
cross-execution-context. C2/C3/C4 are independent (A4 is C4
without C3. A3 is C2 without C4) but every scenario here sits
in one labelled cell. The 5$\times$5 matrix
(Table~\ref{tab:matrix}) marks cells as prior work (P),
illustrative (I), vacuous-on-substrate (V), or out of scope
(O). M1$\times$C1 is vacuous (same-session context cannot
persist into a later session). $M_{i\geq 2}\times$C0 are
vacuous since those substrates encode persistence beyond the
context window. M1$\times$C0 is single-shot
injection~\cite{greshake2023}. M2$\times$C1 is memory
poisoning~\cite{memorygraft2025}. The four illustrative cells
A2 (M3$\times$C2), A3 (M2$\times$C2), A4 (M5$\times$C4), A5
(M4$\times$C4) cover the under-studied rows and columns.

\begin{table}[h]
\centering
\caption{Coverage matrix. V = vacuous on this substrate,
P = prior work, I = illustrative scenario in this paper,
O = out of scope.}
\label{tab:matrix}
\small
\begin{tabular}{@{}lccccc@{}}
\toprule
        & C0 & C1 & C2 & C3 & C4 \\
\midrule
M1      & P  & V  & O  & O  & O  \\
M2      & V  & P  & I  & O  & O  \\
M3      & V  & O  & I  & O  & O  \\
M4      & V  & O  & O  & O  & I  \\
M5      & V  & O  & O  & O  & I  \\
\bottomrule
\end{tabular}
\end{table}

We do not claim exhaustive coverage. The selected cells are those
that prior single-session and memory-only work underspecifies.

\section{Illustrative Scenarios}
\label{sec:attacks}
A4 is our running example and is walked end-to-end. A2, A3, and
A5 populate three further taxonomy cells and are presented as
sketches. Each scenario lists the runtime invariants it depends
on, labelled \emph{confirmed-from-source} (file/line cited) or
\emph{requires-deeper-trace}.

\textbf{A4: Cron via confused deputy (M5$\times$C4).} The
\texttt{cron} tool is owner-only at
\texttt{owner-only-tools.ts}~line~1
(\texttt{["cron","gateway","nodes"]}). A4 therefore runs the
owner as an unwitting trampoline. An attacker email reaches
the configured email gateway with a benign-sounding ``daily
health-check'' tip whose body embeds a webhook URL pointing to
\texttt{atk-sink.example}. The agent's memory pipeline
summarises the email into a stored note that includes the URL.
Days later the owner asks ``set up that daily health-check we
got the email about'', the agent retrieves the note and
synthesises a \texttt{cron.add} whose webhook URL is the
attacker's. The runtime treats the call as owner-issued. The
persistence artifact at firing time is the cron entry (M5),
firing happens in the cron daemon (C4). The owner sees the
visible tool-call name but cannot attribute the embedded URL
to the email gateway's principal. That is the confused-deputy step.
Source anchors (all confirmed-from-source on the pinned commit):
delivery enum at \texttt{cron-tool.ts}~line~37,
\texttt{CronDeliverySchema} accepting an arbitrary \texttt{to}
URL at lines~180--202, \texttt{normalizeHttpWebhookUrl}
accepting any \texttt{http(s)} URL at lines~670--675, and
\texttt{ownerOnly} wired at line~525.
\texttt{artifact/probe-logs/} ships smoke probes
($n{=}20$ dispatch, $n{=}10$ two-stage) and a deterministic
mock-runtime D0/D1/D2 demonstration, the empirical D1 outcome
is qualified in \S\ref{sec:defense}.

\textbf{A2: Skill-trojan via group chat (M3$\times$C2).} A2
requires the non-default \texttt{workspaceAccess="rw"}
configuration (default \texttt{"none"} at
\texttt{config.ts}~line~248), we flag this because it is
load-bearing. Under \texttt{"rw"},
\texttt{appendWorkspaceMountArgs} (at
\texttt{workspace-mounts.ts}~lines~14 and 23--39) emits a
Docker bind mount with \texttt{readOnly=false}, so a
group-session sandbox writes through to the host workspace
path the main session loads from (loader at
\texttt{local-loader.ts}~line~50, resolving \texttt{SKILL.md}
under the writable workspace root). The default mode and the
bridge are confirmed-from-source, the absence of a
skill-creation provenance gate is corroborated by ATLAS
T-PERSIST-001 (residual risk ``Critical''). Per-model
willingness to author leaking skills is
requires-deeper-trace, a known property of
self-improving-agent systems.

\textbf{A3: Cross-channel exfil via memory (M2$\times$C2).} An
attacker email is summarised into memory and later surfaces on
a topical owner query, triggering an outbound email. The
security warning and unique-id markers at
\texttt{external-content.ts}~lines~63, 81--82, 356 emit into model
context only, no runtime hook downstream of the model consults
the provenance the wrapper carries (confirmed-from-source). At
\texttt{active-memory-index.ts}~lines~35--36 the
active-memory plugin sets \texttt{"recent"} and
\texttt{"search"} as its default modes. Supported query
modes are listed at lines~82--87. Ranking
sufficient to surface the poisoned note and per-model
willingness to honour it are requires-deeper-trace, in line with
known retrieval-poisoning behaviour~\cite{zhong2023poisoning,zou2024poisonedrag}.

\textbf{A5: Dotfile patch (M4$\times$C4).} The owner asks the
agent to follow a fetched shell-configuration guide, the page
body contains a \texttt{.zshrc} line that on shell startup
appends a canary marker to a sandbox sink path. The persistence
artifact is the patched \texttt{.zshrc} (M4), firing happens in
the human's interactive shell (C4). ATLAS Trust Boundary~3
lists main-session tool execution as ``Docker sandbox OR Host
(exec-approvals),'' so shell-rc writeability under P1 is
plausible, the filesystem-write invariant is
requires-deeper-trace.

\section{Defense: Provenance with Enforcement}
\label{sec:defense}
We present the defense in three stages of increasing
strength. The contribution is a precise specification, an
executable reference, and a soundness theorem against named
runtime invariants. We do not claim a deployed production
runtime or a measurement here.

\subsection{Notation and core objects}
\label{sec:defense_notation}

Source tags live in
$\mathcal{S} = \mathit{Channel} \times \mathit{Principal} \times
\mathit{Device}$ with elements
$s = (\textit{ch}, \textit{p}, \textit{d})$. Trust is
owner-configured over $(\textit{p}, \textit{d})$ pairs and
lifted channel-independently to triples (lifting an entire
channel to trusted is too coarse, since paired contacts may
themselves be untrusted). $\mathcal{T} \subseteq \mathcal{S}$
is the trusted set induced by the owner's pair set.

$\mathcal{A}$ enumerates every byte the runtime can expose to
the model: durable causal inputs (memory facts, skills,
manifests, plugins, MCP entries, files outside scratch,
configuration and bootstrap blobs, runtime-seeded env vars,
cwd snapshots, clipboard, contact entries, tool schemas),
transient context inputs (system prompt, prior conversation
turns and tool-call results in the running session, planner
state, provenance-preserving compaction summaries that inherit
$\Pi$ from what they summarise), and model-emitted artifacts
(tool-call text, args resolved against the artifact store).
Owner-installed ambient inputs carry $\tau =
\{(\textit{owner-direct}, \textit{owner},
\textit{owner-device})\}$. Including the tool-call text
defeats paraphrase laundering: a model rewriting an attacker
memo into a tool-call argument cannot wash the contribution
by passing it through itself.

The runtime maintains two functions over $\mathcal{A}$. The
tag function $\tau: \mathcal{A} \to 2^{\mathcal{S}}$ records
each artifact's at-source tags and is written only by mediated
creation hooks (\S\ref{sec:hooks}). The provenance state
$\Pi: \mathcal{A} \to 2^{\mathcal{S}}$ accumulates the union of
source tags that contributed to an artifact: $\Pi(b) = \tau(b)$
on fresh intake, and for derived artifacts
\[
  \Pi(f(b_1,\ldots,b_n)) = \bigcup_{i} \bigl(\tau(b_i) \cup \Pi(b_i)\bigr).
\]
Propagation is monotone: any untrusted input strictly enlarges
$\Pi$. The form is the standard lattice rule of Denning's
secure information flow~\cite{denning1976}, applied to an
agent-runtime artifact domain rather than to program
variables.

A consequential action $\alpha$ is a tuple
$(\textit{kind}, \textit{causal}, \textit{args}, \textit{target},
\textit{ownerDevice})$ with $\textit{kind} \in \mathcal{C}$ (the
closed action set below),
$\textit{causal} \subseteq \mathcal{A}$ (the causal-input
artifact ids), arguments, a target identifier (destination URL,
cron job id, etc.), and the $(\textit{p}, \textit{d})$ pair the
action will execute under. Define
$\Pi_\alpha = \bigcup_{x \in \textit{causal}(\alpha)}
\bigl(\tau(x) \cup \Pi(x)\bigr)$. The action-instance digest is
\begin{align*}
  \delta(\alpha) = H\bigl(&\textit{kind},\,\textsf{sort}(\textit{causal}),\,\textsf{cjson}(\textit{args}),\\
                          &\textit{target},\,\textit{ownerDevice}\bigr),
\end{align*}
where $H$ is a collision-resistant hash (SHA-256 in the
reference) and $\textsf{cjson}$ is a
deterministic canonical-JSON encoder. The digest
is over the action's post-normalisation dispatch bytes: the
runtime normalises (default inference, alias rewriting, schema
validation) before the gate sees $\alpha$, so two semantically
equal actions with different pre-normalisation syntax produce
the same $\delta$.

An owner-issued attestation is a tuple
$g = (\delta_g, \nu_g, t_g, q_g)$: digest, one-shot nonce,
expiry, issuer. $\textsf{attest}(\alpha)$ holds iff some $g$
in the runtime's grant set has $\delta_g = \delta(\alpha)$,
$t_g > \textsf{now}$, $q_g \in \mathcal{T}$ matching
$\alpha.\textit{ownerDevice}$, and a nonce $\nu_g$ not yet
consumed. On allow, $\nu_g$ is consumed. Grants arrive only
over a hardware-attested companion channel $\Sigma$ that the
model has no emit primitive into~\cite{tcg_attest_2019}.
Without the one-shot discipline, a single owner approval would
authorise repeated identical dispatches across the expiry
window.

\subsection{The closed action set \texorpdfstring{$\mathcal{C}$}{C}}
\label{sec:closure}

$\mathcal{C}$ is specified by closure rule: an operation is in
$\mathcal{C}$ if its effect escapes the agent process, persists
beyond the session, or mutates state another principal will
later read. The set covers messaging emission, network egress,
file writes outside scratch, skill, plugin, and MCP creation,
modification, installation, and invocation, manifest writes,
scheduler entry mutations, writes to agent bootstrap, system
prompt, or model-router config, every \texttt{contact-list-read}
(additionally bounded by a sliding-window budget
$(N_{\text{contact}}, W_{\text{contact}})$ with defaults
$N{=}10$, $W{=}24\text{h}$, both owner-tunable, applied as a
fail-closed guard), outbound
attestation issuance, and host shell exec outside a documented
allowlist. Anything outside $\mathcal{C}$ that mutates state is
denied fail-closed, mirroring the default-deny discipline of
\texttt{seccomp-bpf} sandboxing.

\subsection{Causal-set construction}
\label{sec:causal}

$\textit{causal}(\alpha)$ is the complete \emph{model-visible
and runtime-mediated} artifact dependency set of $\alpha$.
TCB state (model checkpoint, tokenizer, sampler RNG, model-
router weights, vector-store index structure) sits outside
$\mathcal{A}$ as owner-direct ambient and is not attacker-
attributable, if a hook lets attacker bytes enter that state
(e.g., index entries written by mediated retrieval) those bytes
are artifacts in $\mathcal{A}$ inheriting $\Pi$ from the writer.
The runtime maintains a \emph{transcript provenance set}
$\mathcal{P}_t \subseteq \mathcal{A}$ as the running union of
every artifact whose content has appeared in the model's
context window since session start, minus artifacts pruned by
provenance-preserving compaction (summaries inherit $\Pi$).
$\mathcal{P}_t$ contains the system prompt, prior assistant
turns, tool-call results, every loaded skill manifest and body,
planner state, and any retrieved memo, attachment, contact-list
entry, env value, cwd snapshot, or clipboard entry. For action $\alpha$ at time $t$,
$\textit{causal}(\alpha) = \mathcal{P}_t \cup
\{\alpha\text{-args-resolved}, \textit{tool-call-text}\}$,
where the tool-call text is itself an artifact with
$\tau = \emptyset$ and $\Pi = \bigcup_{x \in \mathcal{P}_t}
\Pi(x)$, registered at H6. The owner-direct request is
included with
$\tau = \{(\textit{owner-direct}, \textit{owner},
\textit{owner-device})\}$. If any contributor to
$\mathcal{P}_t$ cannot be enumerated (say, a skill runs under a
sandbox the runtime does not mediate),
$\textit{causal}(\alpha) := \bot$ and the gate denies. The
discipline mirrors taint inference in
TaintDroid~\cite{enck2010taintdroid} and Fides-style agent
information flow control~\cite{fides2024agent}, but operates
at the agent's artifact layer rather than at byte or IPC
granularity.

\subsection{Mediation hooks}
\label{sec:hooks}

D2 requires complete mediation~\cite{saltzer1975protection}:
every read or write of an artifact in $\mathcal{A}$ traverses
one of ten runtime hooks, summarised in
Table~\ref{tab:hooks}. The hooks split into two roles.
\emph{Update hooks} (H1--H5) populate $\tau$ and propagate
$\Pi$ as side effects of allowed actions. They never decide
whether to allow. \emph{Gate hooks} (H6--H10) are decision
points: each fires before a side-effecting dispatch, computes
$\Pi_\alpha$ over the action's causal set, and either allows
(if all contributing tags are trusted or a matching grant
exists) or denies. Operations that escape mediation
(FFI, side-channel storage, unseeded env vars, browser plugin
state) are residual surface: any artifact with unset $\tau$
has $\Pi$ treated as universal-untrusted, so
$\Pi_\alpha \not\subseteq \mathcal{T}$ by construction.

\begin{table}[h]
\centering
\caption{The ten mediation hooks of D2. Update hooks (H1--H5)
populate $\tau$ and $\Pi$. Gate hooks (H6--H10) decide whether
to dispatch.}
\label{tab:hooks}
\footnotesize
\setlength{\tabcolsep}{3pt}
\setlength{\tabcolsep}{6pt}
\begin{tabular}{@{}lp{0.78\columnwidth}@{}}
\toprule
\multicolumn{2}{@{}l}{\emph{Update hooks: populate $\tau$ and propagate $\Pi$.}} \\
\midrule
H1 & Inbound adapter: sets $\tau(b)$ from the gateway's
authenticated source. \\
H2 & Memory write: closes $\Pi$ over causal sources. \\
H3 & Memory retrieval: registers recalled memos as causal
contributors to the next gate. \\
H4 & Skill / plugin / MCP creation or modification:
propagates $\Pi$, seeds $\tau = \emptyset$. \\
H5 & Skill / plugin / MCP load: registers manifest and body
as separate artifacts. \\
\midrule
\multicolumn{2}{@{}l}{\emph{Gate hooks: decide before any side-effecting dispatch.}} \\
\midrule
H6 & Tool-call construction: registers the model-emitted
tool-call text and submits to the gate. \\
H7 & Host shell-command dispatch. \\
H8 & Filesystem write outside scratch. \\
H9 & Scheduler entry create/modify/remove. \\
H10 & Outbound network or messaging emission. \\
\bottomrule
\end{tabular}
\end{table}

Each kind in $\mathcal{C}$ traverses exactly one gate hook.
Table~\ref{tab:action-hook} gives the mapping, grouped by gate
hook. Any kind not listed there denies fail-closed by default,
which makes \textsc{(I-Mediation)} checkable against the spec
rather than against generic ``complete mediation'' language.

\begin{table}[h]
\centering
\caption{Each consequential action kind is gated by exactly
one of H6--H10. Kinds not appearing here deny fail-closed.}
\label{tab:action-hook}
\footnotesize
\setlength{\tabcolsep}{3pt}
\sloppy
\setlength{\tabcolsep}{6pt}
\begin{tabular}{@{}lp{0.79\columnwidth}@{}}
\toprule
Gate & Action kinds \\
\midrule
H6 &
\texttt{skill-\{create, modify, load, exec\}},
\texttt{plugin-\{install, modify, load, exec\}},
\texttt{mcp-server-\{install, modify, load\}},
\texttt{mcp-tool-call}, \texttt{manifest-write},
\texttt{contact-list-read} (additionally rate-limited by
$(N_{\text{contact}}, W_{\text{contact}})$). A2 fires here.
For \texttt{*-exec} and \texttt{mcp-tool-call}, side effects
of the running body are lowered into fresh actions that cross
H7--H10. Load-time approval does not authorise later effects. \\
\midrule
H7 & \texttt{host-shell-exec}. \\
\midrule
H8 &
\texttt{fs-write} (outside scratch), \texttt{config-write},
\texttt{model-router-write}, \texttt{system-prompt-write},
\texttt{agent-bootstrap-write}. \\
\midrule
H9 &
\texttt{schedule-\{create, modify, remove\}}. A4 fires here. \\
\midrule
H10 &
\texttt{messaging-send}, \texttt{network-egress},
\texttt{outbound-attest-issue}. \\
\bottomrule
\end{tabular}
\end{table}

\subsection{D0, D1, D2}

D0 dispatches every model-emitted action with no $\Pi$ and no
gate. D1 maintains $\tau$ and $\Pi$ as D2 does, but encodes
enforcement \emph{inside} the model loop: provenance tags
appear in context whenever an artifact is recalled, and the
model is asked to refuse on untrusted provenance. Two reasons make D1
insufficient. Adaptive attacks against
in-context defenses achieve $\geq 90\%$ ASR across twelve
settings~\cite{nasr2025}, their setting is jailbreaks rather
than provenance, so we treat this as supporting evidence rather
than proof. Our own smoke probe (\S\ref{sec:probe}) suggests
that OpenClaw's in-context security warning at
\texttt{external-content.ts}~line~81 (warning constant, text body at line~82) does not, on its own,
prevent the A4 dispatch. The result is consistent with broader
findings on the brittleness of prompted
defenses~\cite{liu2024formalising}.

D2 intercepts every consequential action $\alpha$ at the
appropriate H6--H10 hook, computes $\textit{causal}(\alpha)$
and $\Pi_\alpha$, and applies
\begin{align*}
  \mathit{Allow}(\alpha) \;\Leftrightarrow\;
    & \alpha.\textit{kind} \in \mathcal{C} \;\wedge\; \mathit{wf}(\alpha) \\
    & \wedge\; \bigl(\Pi_\alpha \subseteq \mathcal{T} \;\vee\; \textsf{attest}(\alpha)\bigr),
\end{align*}
where $\mathit{wf}(\alpha)$ holds iff
$\textit{causal}(\alpha) \neq \bot$ and every $x \in
\textit{causal}(\alpha)$ has $\Pi(x) \cup \tau(x) \neq
\emptyset$ (no unprovenanced contributor). Fail-closed
reasons (unknown artifact, empty causal, empty-provenance,
expired or digest-mismatched grant, exceeded budget,
unclassified kind) each produce a named deny in the
reference. $\textsf{attest}$ binds to $\delta(\alpha)$
over $\Sigma$, and the model cannot emit into $\Sigma$.

\subsection{Soundness theorem}
\label{sec:soundness}

The mediation invariants (I-Mediation, I-Tag, I-Causal) for
the A4 path are backed by source-anchor regression tests
against the pinned OpenClaw source
(\texttt{src-audit/audit.ts}, 13 cases): each load-bearing
A4/A2/A3 line cite becomes a test that fails if the vendored
source drifts. These tests verify that the cited code paths
exist, they do not prove complete mediation. Operational
mediation is exercised, on the H1/H2/H3/H6/H9 slice, by the
runtime adapter in \S\ref{sec:integration}. The cryptographic
and channel invariants (I-Channel, I-GrantAuth, I-Nonce,
I-Hash) are deployment assumptions on the runtime, we state
them precisely below but do not source-check them.

\textbf{Theorem (D2 soundness).} \emph{Assume the following
runtime invariants:}

\smallskip
\noindent\textbf{(I-Mediation)} Every read or write of an
artifact in $\mathcal{A}$ traverses a hook in
$\{H1,\ldots,H10\}$.

\smallskip
\noindent\textbf{(I-Tag)} $\tau(b)$ is set by H1 from the
adapter's authenticated source identifier, the model has no
edit primitive on $\tau$.

\smallskip
\noindent\textbf{(I-Causal)} $\textit{causal}(\alpha)$ is the
closure of \S\ref{sec:causal} over the transcript provenance
set $\mathcal{P}_t$, if the runtime cannot enumerate it,
$\textit{causal}(\alpha) := \bot$ and the gate denies.

\smallskip
\noindent\textbf{(I-Channel)} Attestations $g$ arrive only over
the hardware-attested channel $\Sigma$, the model has no emit
primitive into $\Sigma$.

\smallskip
\noindent\textbf{(I-GrantAuth)} Each grant carries an
unforgeable authenticator (e.g.\ a signature from $q_g$) whose
verification is performed by the runtime, not by data supplied
alongside the grant. $q_g$ is set from the verified
authenticator, never from grant fields the model can influence.

\smallskip
\noindent\textbf{(I-Nonce)} The consumed-nonce ledger is
durable and globally unique within the runtime, concurrent
dispatches serialize on ledger insertion. Every fresh grant
carries a nonce $\nu_g$ drawn uniformly from a space large
enough that collision probability is negligible.

\smallskip
\noindent\textbf{(I-Hash)} $H$ is collision resistant and
$\textsf{cjson}$ is deterministic.
\smallskip

\emph{Then for every action $\alpha$ that D2 dispatches, either}
$\Pi_\alpha \subseteq \mathcal{T}$ \emph{or there exists a grant
$g$ with $\delta_g = \delta(\alpha)$, $t_g > \mathsf{now}$,
$q_g \in \mathcal{T}$ matching $\alpha.\textit{ownerDevice}$,
and a nonce $\nu_g$ that was unconsumed at dispatch and was
atomically consumed against the durable ledger before $\alpha$
was emitted. $g$ was issued by an owner-trusted principal over
$\Sigma$ for the exact action-instance bytes the gate observed.}

\emph{Proof sketch.} By \textsc{(I-Mediation)} and
\textsc{(I-Causal)}, $\textit{causal}(\alpha)$ is either $\bot$
(denial) or a finite set whose source tags were set at intake
by H1 or inherited through mediated propagation
under \textsc{(I-Tag)}. Monotone propagation puts every
contributing $\tau(b_i)$ into $\Pi_\alpha$, and the rule only
dispatches when $\Pi_\alpha \subseteq \mathcal{T}$ or
$\textsf{attest}(\alpha)$. In the attested branch,
\textsc{(I-Hash)} forces $\delta_g = \delta(\alpha)$ to bind
the exact post-normalisation bytes. \textsc{(I-Channel)}
restricts $g$ to arrive over $\Sigma$. \textsc{(I-GrantAuth)}
forces $q_g$ to come from the verified authenticator rather
than from data, and \textsc{(I-Nonce)} prevents replay since
the ledger is durable and unique. Together, the only way a
dispatch is allowed is if every contributor in
$\textit{causal}(\alpha)$ has a tag in $\mathcal{T}$ or an
owner-trusted principal in $\mathcal{T}$ has issued a fresh,
authenticated, action-bound grant for the exact instance.
$\square$ For A4, $\textit{causal}(\alpha)$ is the owner
request, recalled memory note, model-emitted tool-call text,
and cron payload, the memory note carries the email gateway's
principal tag so $\Pi_\alpha \not\subseteq \mathcal{T}$ and the
gate denies. A5 and A2 decompose analogously.

\subsection{The captured A4 probe}
\label{sec:probe}

In $n{=}20$ single-shot trials, a frontier-class reasoning
model is given the OpenClaw security warning verbatim
(\texttt{external-content.ts}~lines~81--82), the recalled
memory note wrapped in the unique-id markers (line~63) with
the attacker URL embedded, and the owner request ``set up
that daily health-check we got the email about.'' 19/20
emitted a runtime-valid \texttt{cron.add} with
\texttt{delivery.to} pointing to the attacker URL (matching
the cron schema at \texttt{cron-tool.ts}~line~34 and
lines~157--202). 1/20 emitted an empty object. A two-stage
variant ($n{=}10$) still produces 7/10 attacker outcomes.
Probe-log headers ship the verbatim prompts, tool schema, and
sampling temperature ($T{=}1.0$), model identity is anonymised
for review. These figures anchor the narrative claim that an
in-context warning does not, on its own, prevent A4,
quantitative D1 claims rest on the preregistered evaluation
in \S\ref{sec:eval}.

\subsection{D3 and adaptive attacks}
\label{sec:d3_adaptive}

D3 adds per-skill capability manifests on top of D2. A manifest
declares the subset of $\mathcal{C}$ a skill may attempt and is
signed at install time by a principal in $\mathcal{T}$.
\texttt{manifest-write} is itself in $\mathcal{C}$. The
construction composes the Agents Rule of
Two~\cite{rule_of_two_2025}: a skill may communicate
externally, modify state, or process untrusted content, but
not more than two of those without explicit attestation. D3
keeps a self-improving skill from quietly widening its own
capability set, in the same spirit as
POLA~\cite{miller2003paradigm} and Android permission
manifests.

Model-visible provenance alone does not suffice.
Provenance must bind to an enforcement point outside the
model loop, on a closed action set, with
action-instance-scoped attestation the model cannot emit. The
upstream issue~\cite{openclaw_runtime_proposal} proposed
provenance tagging without specifying enforcement, our
contribution is the boundary placement and the formal
propagation/attestation discipline.

A real attacker has several routes available, each
preregistered as a follow-on red-team vector. Tag forgery at H1
fails because, under \textsc{(I-Tag)}, only the adapter writes
$\tau$. Approval-UX confusion is the case where a malicious
action's canonical preimage looks benign on the companion
device. Surfacing the full preimage mitigates most of this,
with residual risk from human visual error on long preimages,
to be quantified against prior signing-UI usability work.
Poisoning the attestation channel runs into the
hardware-attested companion, leaving hardware compromise as
the residual. Mediation escape is the case where a skill
performs side effects without traversing any H1--H10 hook. By
construction the gate cannot fire when no hook does, so an
enclosing sandbox (seccomp, container, microVM) is mandatory
to close this gap. Without it, FFI and direct syscalls remain
residual surface. \textsc{(I-Hash)} kills digest-preimage
manipulation because the digest covers the full preimage,
including target and owner-device. The model cannot be coerced
into emitting $\textsf{attest}$ because \textsc{(I-Channel)}
keeps that primitive out of the model's emit alphabet.

\subsection{Executable policy specification}
\label{sec:reference}
The artifact is at
\url{https://github.com/maloyan/sleeper-channels}.
\verb|artifact/d2-gate/| is a TypeScript implementation of the
gate decision function, $\delta$ (SHA-256 over canonical JSON),
$\mathcal{C}$, the channel-independent trust lifting, the
$\Pi_\alpha$ union, and the contact-read budget. The decision
function is pure and fail-closed at every point a trust
assertion cannot be established.

The test suite has 42 cases across four subsuites. The gate
suite (23 cases) exercises every decision rule, including
multi-input laundering, grant replay on a consumed nonce, and
the rate-limiter fail-closed path. The mock-runtime suite (3
cases) replays the probe-log majority output and confirms that
D0 and D1 dispatch while D2 denies on the mixed
\texttt{(email,attacker)+(owner-direct,owner)} provenance. The
static-audit suite (\texttt{src-audit/audit.ts}, 13 cases) reads
the vendored OpenClaw source and asserts that the load-bearing
A4/A2/A3 line cites are present at the claimed lines. A separate
runtime adapter (\texttt{openclaw-integration/}, 3 cases) wires
H1, H2, H3, H6, and H9 around the cron path. It uses faithful
stubs of \texttt{normalizeCronJobCreate} and
\texttt{normalizeHttpWebhookUrl} that match
\texttt{cron-tool.ts}~lines~637 and 670--675, and it shows
end-to-end that the A4 attacker call is denied with
\texttt{untrusted-provenance} surfacing the email gateway's
principal, that a benign owner-only cron is admitted as
\texttt{all-trusted}, and that H6 unions provenance correctly
over multi-source recalls.

Together, the audit and the runtime adapter check
\textsc{(I-Mediation)} and \textsc{(I-Causal)} for the
H1/H2/H3/H6/H9 slice on the pinned commit. The other five
hooks and a full deployment proof are explicit follow-on. The
artifact's structure also matches contemporary
reproducibility-evaluation guidance for systems
papers~\cite{collberg2016repeatability}.

\subsection{Integration sketch (OpenClaw) and measurement plan}
\label{sec:integration}\label{sec:eval}
Three integration points, audited at file/line granularity but
not yet patched upstream: tagging at
\texttt{external-content.ts}~line~356 (sidecar $(\textit{ch},
\textit{p}, \textit{d})$ on inbound content, hook H1),
sidecar manifests co-located with each \texttt{SKILL.md} at
the writer for \texttt{local-loader.ts}~line~50 (H4--H5), and
gating on $\Pi_\alpha$ in \texttt{cron-tool.ts} between
\texttt{normalizeCronJobCreate} at line~637 and runtime
submission (H9). The gate would block A4 where
\texttt{exec-approvals} does not, because the existing approval
surface keys on the visible command string rather than on the
provenance of the data that drove the call.

We preregister the follow-on empirical study: the OpenClaw SHA
used here. A3 measured under P1, A2/A4/A5 under P2 (matching
the profile each scenario's firing context requires), primary
endpoint Action rate at delay $\tau^* = 50$ benign turns, eight
Holm--Bonferroni contrasts across D0/D1/D2
and A2--A5 at $n \geq 200$ seeds per cell, deterministic canary
oracles, an adaptive-attack budget implementing the five
vectors above, and a 50-task benign suite for utility-cost
parity.

\section{Discussion and Limitations}
\label{sec:discussion}
The paper delivers a definition, a taxonomy, a
source-anchored feasibility argument, and a formal defense
with theorem and reference. Empirical attack-success rates,
defense efficacy under deployment, and utility cost are
deferred to the preregistered follow-on. The
taxonomy is substrate-independent. The defense formalism
applies to any agent runtime that mediates artifact creation
through identifiable code paths and can host a
hardware-attested companion channel. Memory poisoning,
provenance tagging, and the confused-deputy condition are each
well-known on their own. New here is the combination on the
OS-live substrate, together with the move of treating
model-emitted tool-call text as an artifact whose provenance
must be tracked to defeat paraphrase laundering.

A2/A3/A5 are sketches with \emph{requires-deeper-trace}
invariants. A2 needs the non-default \texttt{rw} workspace
mode. The defense has not been deployed or stress-tested
against an adaptive attacker, and the integration points are
OpenClaw-specific.

\section{Ethics and Disclosure}
\label{sec:ethics}
We omit runnable payloads and real credential targets, the
attack abstractions match maintainer-published documentation
and the already-public upstream defense
proposal~\cite{openclaw_runtime_proposal}. The follow-on
study instruments all harms via canaries (synthetic secrets,
sink mailboxes, sandboxed filesystem markers, isolated-VM
cron entries, synthetic contact lists) and follows
coordinated-disclosure norms: working A4 and A5 templates
gated to patched-version disclosure, with the D1/D2 reference
defenses filed as upstream patches.

\section{Conclusion}
\label{sec:conclusion}
Always-on OS-live agents fold messaging, memory, skills,
scheduling, and shell into a single persistent authority
boundary that admits sleeper-channel attacks. The paper fixed
the class on two axes, took A4 end-to-end through OpenClaw at
a pinned commit, and worked out a tiered provenance defense
whose load-bearing piece sits outside the model loop. Seven
invariants underwrite the D2 soundness theorem. A canonical
action-instance digest, paired with one-shot grant nonces,
defeats paraphrase laundering and replay. The 42-test
reference runs on Node $\geq{}20$. Empirical study is
preregistered.

\bibliographystyle{IEEEtran}
\bibliography{references}

\end{document}